\documentclass[12pt]{article}
\usepackage{epsfig}
\def\be{\begin{equation}}
\def\ee{\end{equation}}
\def\bea{\begin{eqnarray}}
\def\eea{\end{eqnarray}}
\usepackage{graphicx}% Include figure files

\catcode`\@=11
\def\lsim{\mathrel{\mathpalette\@versim<}}
\def\gsim{\mathrel{\mathpalette\@versim>}}
\def\@versim#1#2{\vcenter{\offinterlineskip
\ialign{$\m@th#1\hfil##\hfil$\crcr#2\crcr\sim\crcr } }}
\catcode`\@=12
\usepackage{axodraw}

\catcode`@=12
\begin{document}
\thispagestyle{empty}
\begin{flushright}
UCRHEP-T489\\
April 2010\
\end{flushright}
\vspace{1.0in}
\begin{center}
{\LARGE \bf Utilitarian Supersymmetric Gauge Model\\
of Particle Interactions\\}
\vspace{1.2in}
{\bf Ernest Ma\\}
\vspace{0.2in}
{\sl Department of Physics and Astronomy, University of California,\\
Riverside, California 92521, USA\\}
\end{center}
\vspace{1.2in}
\begin{abstract}\
A remarkabale U(1) gauge extension of the supersymmetric standard model was 
proposed eight years ago.  It is anomaly-free, has no $\mu$ term, and conserves 
baryon and lepton numbers automatically.  The phenomenology of a specific 
version of this model is discussed.  In particular, leptoquarks are predicted, 
with couplings to the heavy singlet neutrinos, the scalar partners of which may 
be components of dark matter.  The Majorana neutrino mass matrix itself may 
have two zero subdeterminants.
\end{abstract}

\newpage
\baselineskip 24pt

\noindent \underline{\it Introduction}~:~ The transition from the Standard 
Model (SM) of particle interactions to the Minimal Supersymmetric Standard 
Model (MSSM) is fraught with two well-known shortcomings. (1) Whereas baryon 
number $B$ and lepton number $L$ are automatically conserved in the SM, they 
are conserved in the MSSM only by the imposition of $R$ parity, i.e. $R \equiv 
(-1)^{2j+3B+L}$. (2) There exists the term $\mu \phi_1 \phi_2$  in the MSSM 
superpotential, where $\phi_{1,2}$ are the two Higgs superfields which 
spontaneously break the electroweak gauge symmetry.  Since this term is 
allowed by the gauge symmetry and the supersymmetry, there is no understanding 
of why $\mu$ should be of order the electroweak breaking scale, rather 
than some very large unification scale.

It is clearly desirable and useful to have a single mechanism which solves 
both problems.  One such utilitarian proposal was made eight years 
ago~\cite{m02}, using a new $U(1)_X$ gauge symmetry.  Let the quark and 
doublet superfields transform as $n_1$ and $n_4$ respectively under $U(1)_X$.  
Requiring the absence of anomalies, two classes of solutions for the other 
superfield assignments are then obtained as functions of $n_1$ and $n_4$.  
In this paper, the particularly simple choice of $n_1=0$ in Solution (A) 
is discussed, together with some of its phenomenology, relevant to the 
operating Large Hadron Collider (LHC).

\noindent \underline{\it Model}~:~ Consider the gauge group $SU(3)_C \times
SU(2)_L \times U(1)_Y \times U(1)_X$ with the particle content of 
Ref.~\cite{m02}.  For $n_1=0$ in Solution (A), the various superfields 
transform as shown in Table 1.  There are three copies of 
$Q,u^c,d^c,L,e^c,N^c,S_1,S_2$; two copies of $U,U^c,S_3$; and one copy of 
$\phi_1,\phi_2,D,D^c$.
\begin{table}[htb]
\caption{Particle content of proposed model.}
\begin{center}
\begin{tabular}{|c|c|c|c|c|}
\hline
Superfield & $SU(3)_C$ & $SU(2)_L$ & $U(1)_Y$ & $U(1)_X$ \\
\hline
$Q = (u,d)$ & 3 & 2 & 1/6 & 0 \\
$u^c$ & $3^*$ & 1 & $-2/3$ & 3/2 \\
$d^c$ & $3^*$ & 1 & 1/3 & 3/2 \\
\hline
$L = (\nu,e)$ & 1 & 2 & $-1/2$ & 1 \\
$e^c$ & 1 & 1 & 1 & 1/2 \\
$N^c$ & 1 & 1 & 0 & 1/2 \\  
\hline
$\phi_1$ & 1 & 2 & $-1/2$ & $-3/2$ \\
$\phi_2$ & 1 & 2 & 1/2 & $-3/2$ \\
$S_1$ & 1 & 1 & 0 & $-1$ \\
$S_2$ & 1 & 1 & 0 & $-2$ \\
$S_3$ & 1 & 1 & 0 & 3 \\
\hline
$U$ & 3 & 1 & 2/3 & $-2$ \\
$D$ & 3 & 1 & $-1/3$ & $-2$ \\
$U^c$ & $3^*$ & 1 & $-2/3$ & $-1$ \\
$D^c$ & $3^*$ & 1 & 1/3 & $-1$ \\
\hline
\end{tabular}
\end{center}
\end{table}
The only allowed terms of the superpotential are thus trilinear, i.e.
\begin{eqnarray}
&& Q u^c \phi_2, ~~~ Q d^c \phi_1, ~~~ L e^c \phi_1, ~~~ L N^c \phi_2, ~~~ 
S_3 \phi_1 \phi_2, ~~~ N^c N^c S_1, \\ 
&& S_3 U U^c, ~~~ S_3 D D^c, ~~~ u^c N^c U, ~~~ u^c e^c D, ~~~ d^c N^c D, ~~~ 
Q L D^c, ~~~ S_1 S_2 S_3.
\end{eqnarray}
The absence of any bilinear term means that all masses come from soft 
supersymmetry breaking, thus explaining why the $U(1)_X$ and electroweak 
symmetry breaking scales are not far from that of supersymmetry breaking. 
As $S_{1,2,3}$ acquire nonzero vacuum expectation values (VEVs), the exotic 
$(U,U^c)$ and $(D,D^c)$ fermions obtain Dirac masses from $\langle S_3 
\rangle$, which also generates the $\mu$ term.  The singlet $N^c$ fermion 
gets a large Majorana mass from $\langle S_1 \rangle$, so that the neutrino 
$\nu$ gets a small seesaw mass in the usual way. The singlet $S_{1,2,3}$ 
fermions themselves get Majorana masses from their scalar counterparts 
$\langle S_{1,2,3} \rangle$ through the $S_1 S_2 S_3$ terms.  The only 
massless fields left are the usual quarks and leptons. They then become 
massive as $\phi^0_{1,2}$ acquire VEVs, as in the MSSM.

Because of $U(1)_X$, the structure of the superpotential conserves both 
$B$ and $(-1)^L$, with $B=1/3$ for $Q,U,D$, and $B=-1/3$ for $u^c,d^c,U^c,D^c$; 
$(-1)^L$ odd for $L,e^c,N^c,U,U^c,D,D^c$, and even for all others. Hence 
the exotic $U,U^c,D,D^c$ scalars are leptoquarks and decay into ordinary quarks 
and leptons.  The $R$ parity of the MSSM is defined here in the same way, 
i.e. $R \equiv (-)^{2j+3B+L}$, and is conserved.  Note also that the 
quadrilinear terms $QQQL$ and $u^c u^c d^c e^c$ (allowed in the MSSM) as 
well as $u^c d^c d^c N^c$ are forbidden by $U(1)_X$.  Proton decay is thus 
strongly suppressed.  It may proceed through the quintilinear term 
$QQQL S_1$ as the $S_1$ fields acquire VEVs, but this is a dimension-six 
term in the effective Lagrangian, which is suppressed by two powers 
of a very large mass, say the Planck mass, and may safely be allowed.

\noindent \underline{\it Gauge sector}~:~ The new $Z_X$ gauge boson of this 
model becomes massive through $\langle S_{1,2,3} \rangle = u_{1,2,3}$, whereas 
$\langle \phi^0_{1,2} \rangle = v_{1,2}$ contribute to both $Z$ and $Z_X$. 
The resulting $2 \times 2$ mass-squared matrix is given by~\cite{km97}
\begin{equation}
{\cal M}^2_{Z,Z_X} = \pmatrix{(1/2)g_Z^2(v_1^2+v_2^2) & (3/2)g_Z g_X 
(v_2^2-v_1^2) \cr (3/2)g_Z g_X (v_2^2-v_1^2) & 2g_X^2 [u_1^2 + 4 u_2^2 + 
9 u_3^2 + (9/4)(v_1^2 + v_2^2)]}.
\end{equation}
Since precision electroweak measurements require $Z-Z_X$ mixing to be very 
small~\cite{elmp09}, $v_1 = v_2$, i.e. $\tan \beta = 1$, is assumed from 
now on.

Consider the decay of $Z_X$ to the usual quarks and leptons.  Each fermionic 
partial width is given by
\begin{equation}
\Gamma(Z_X \to \bar{f} f) = {g_X^2 M_{Z_X} \over 24 \pi} [c_L^2 + c_R^2],
\end{equation}
where $c_{L,R}$ can be read off under $U(1)_X$ from Table 1.  Thus
\begin{equation}
{\Gamma(Z_X \to \bar{t} t) \over \Gamma(Z_X \to \mu^+ \mu^-)} = 
{\Gamma(Z_X \to \bar{b} b) \over \Gamma(Z_X \to \mu^+ \mu^-)} = {27 \over 5}.
\end{equation}
This will serve to distinguish it from other $Z'$ models~\cite{gm08}.

\noindent \underline{\it Effective two-Higgs-doublet structure}~:~ 
In the MSSM, the scalar potential of the two Higgs doublets is given by
\begin{eqnarray}
V &=& m_1^2 \Phi_1^\dagger \Phi_1 + m_2^2 \Phi_2^\dagger \Phi_2 + m_{12}^2 
(\Phi_1^\dagger \Phi_2 + \Phi_2^\dagger \Phi_1) \nonumber \\ 
&+& {1 \over 2} \lambda_1 (\Phi_1^\dagger \Phi_1)^2 + 
{1 \over 2} \lambda_2 (\Phi_2^\dagger \Phi_2)^2 + 
\lambda_3 (\Phi_1^\dagger \Phi_1)(\Phi_2^\dagger \Phi_2) + 
\lambda_4 (\Phi_1^\dagger \Phi_2)(\Phi_2^\dagger \Phi_1),
\end{eqnarray}
where
\begin{eqnarray}
\lambda_{1,2} = {1 \over 4} (g_1^2 + g_2^2), ~~~ 
\lambda_3 = {1 \over 4} (-g_1^2 + g_2^2), ~~~
\lambda_4 = -{1 \over 2} g_2^2.
\end{eqnarray}
In the present model, there are extra Higgs singlets, but if they are heavier 
than the doublets by an order of magnitude and the soft $A$ terms are of the 
elctroweak scale, they can be integrated out and the effective 
two-Higgs-doublet structure is given by~\cite{km97,mn94,dm93,dm95,km96,lm97}
\begin{eqnarray}
\lambda_{1,2} &=& {1 \over 4} (g_1^2 + g_2^2) + f^2 - {f^4 \over 9 g_X^2}, \\ 
\lambda_3 &=& {1 \over 4} (-g_1^2 + g_2^2) + f^2 - {f^4 \over 9 g_X^2}, \\ 
\lambda_4 &=& -{1 \over 2} g_2^2 + f^2,
\end{eqnarray}
where $f$ is the Yukawa coupling of the trilinear term $S_3 \phi_1 \phi_2$, 
assuming that this one particular singlet dominates over all others in 
changing the scalar quartic couplings $\lambda_i$.

Since $\tan \beta = 1$ in this model, the lightest neutral Higgs boson 
has the upper bound~\cite{km97}
\begin{equation}
(m_h^2)_{max} = \epsilon + {f^2 \over \sqrt{2} G_F} \left[ {3 \over 2} - 
{f^2 \over 9 g_X^2} \right],
\end{equation}
where
\begin{equation}
\epsilon = {3 g_2^2 m_t^4 \over 8 \pi^2 M_W^2} \ln \left( 1 + {\tilde{m}^2 
\over m_t^2} \right)
\end{equation}
is the well-known large radiative correction~\cite{oyy91,hh91,erz91,bfc91,y91} 
due to the $t$ quark and its supersymmetric scalar partners.  The above 
upper bound may easily exceed that of the MSSM. For example, let $f = 
3 g_X = g_2$, then $(m_h)_{max} = 2M_W^2 + \epsilon = 143$ GeV, assuming 
$\tilde{m} = 1$ TeV in Eq.~(12).  Contrast this with the upper bound in the 
MSSM, i.e. $M_Z^2 \cos^2 2 \beta + \epsilon < (126~{\rm GeV})^2$.  If the 
experimental bound of $m_h > 114.4$ GeV is used, then $|\cos 2 \beta| > 0.81$ 
is required.  Here the prediction is that $\cos 2 \beta = 0$ and yet $m_h$ 
may be substantially greater than 114.4 GeV.   This difference will have 
important implications for the Higgs search at the LHC.  Another important 
difference is the mass of the charged Higgs boson:
\begin{equation}
m_{H^\pm}^2 = m_A^2 + M_W^2 - f v^2,
\end{equation}
where $m_A$ is the mass of the pseudoscalar Higgs boson, i.e. $Im (\phi_1^0 - 
\phi_2^0)$.  For example, if $f = g_2$, then $m_{H^\pm}^2 - m_A^2 = -M_W^2$, 
instead of $+M_W^2$ in the MSSM.

\noindent \underline{\it Neutrino masses}~:~  There are three copies each of 
the superfields $L$, $e^c$, $N^c$, and $S_1$.  As such, a family symmetry in 
the lepton sector may be supported.  For example, the discrete symmetry 
$Z_4$ may be used to realize the interesting proposal of Ref.~\cite{m05-1} 
that the observed neutrino mass matrix has two zero subdeterminants.  Under 
$Z_4$, the three copies of $L$, $e^c$, $N^c$, and $S_1$ separately all 
transform as 1, $i$, $-i$, with $\phi_{1,2}$ transforming as 1.  From the 
$L e^c \phi_1$ and $L N^c \phi_2$ couplings, the charged-lepton and Dirac 
neutrino mass matrices are thus diagonal, whereas the $N^c N^c S_1$ couplings 
result in the Majorana mass matrix of the form~\cite{fkmt05,m05-2}
\begin{equation}
{\cal M}_N = \pmatrix{A & B & C \cr B & 0 & D \cr C & D & 0}.
\end{equation}
The resulting seesaw neutrino mass matrix is then given by~\cite{m05-1,l05}
\begin{equation}
{\cal M}_\nu = \pmatrix{\alpha & \beta & \gamma \cr \beta & \alpha^{-1} 
\beta^2 & \delta \cr \gamma & \delta & \alpha^{-1} \gamma^2},
\end{equation}
where the subdeterminants of the $(1,2)$ and $(1,3)$ blocks are clearly zero.
Assuming $\gamma = \beta$, then $\theta_{23} = \pi/4$, $\theta_{13}=0$, 
and
\begin{eqnarray}
\Delta m^2_{21} \cos 2 \theta_{12} &=& |\alpha^{-1} \beta^2 + \delta|^2 
- |\alpha|^2, \\ 
\Delta m^2_{21} \sin 2 \theta_{12} &=& 2 \sqrt{2} |\alpha^* \beta + \beta^* 
(\alpha^{-1} \beta^2 + \delta)|, \\ 
\Delta m^2_{32} &=& |\alpha^{-1} \beta^2 - \delta|^2 - {1 \over 2} |\alpha|^2 
- {1 \over 2} |\alpha^{-1} \beta^2 + \delta|^2 - 2 |\beta|^2.
\end{eqnarray}
Since there can only be one nontrivial phase in the above, let $\alpha$ be 
real, $\epsilon = \alpha^{-1} \beta^2 + \delta$ real, and $\beta$ 
complex.  In the case $\beta$ is also real, it has already been 
shown~\cite{m05-1} that the resulting solution has the normal 
hierarchy of neutrino masses, i.e. $m_1 < m_2 < m_3$, such that the 
effective neutrino Majorana mass measured in neutrinoless double beta decay, 
i.e. the parameter $\alpha$, is about $6 \times 10^{-4}$ eV, much below the 
sensitivity of such experiments.

Consider now the case of a purely imaginary $\beta$, 
using $\beta = i \zeta$, where $\zeta$ is real, then a solution exists 
where $\alpha$ is several times larger, as shown below.  Rewriting Eqs.~(16) 
to (18),
\begin{eqnarray}
\Delta m^2_{21} \cos 2 \theta_{12} &=& (\epsilon - \alpha)(\epsilon + \alpha), 
\\ 
\Delta m^2_{21} \sin 2 \theta_{12} &=& 2 \sqrt{2} \zeta (\epsilon - \alpha), 
\\ 
\Delta m^2_{32} &=&  \zeta^2 (4\alpha^{-2} \zeta^2 + 4 \alpha^{-1} \epsilon  
- 2) + {1 \over 2} (\epsilon^2 - \alpha^2).
\end{eqnarray}
Let $\alpha = 0.0042$ eV, $\epsilon = 0.0067$ eV, and $\zeta = 0.0098$ eV, 
then $\Delta m^2_{21} = 7.45 \times 10^{-5}$ eV$^2$, $\tan^2 \theta_{12} = 
0.464$, and $\Delta m^2_{32} = 2.53 \times 10^{-3}$ eV$^2$, in good agreement 
with data~\cite{gms10}.

\noindent \underline{\it Heavy neutrino singlet at the LHC}~:~
In the canonical seesaw, even if the the heavy neutrino singlet anchor 
has a mass of order TeV, it is very hard to produce at the LHC~\cite{aa09}, 
because it couples only to leptons and the strength of that coupling is 
necessarily very weak, as required by the tiny neutrino mass~\cite{m09}.  
Here, since $N^c$ also couples to the leptoquarks $(U,U^c)$ and $(D,D^c)$, 
it can be produced at the LHC as the decay product of the latter, which 
are copiously produced themselves because they have strong interactions.

Since the scalar $\tilde{N}^c$ has odd $R$ parity, it may also be a component 
of dark matter.  In that case, the decays of the heavy leptoquark fermions 
$(U,U^c)$ and $(D,D^c)$ to quark jets and $\tilde{N}^c$ may lead to sizable 
missing-energy signals at the LHC, as recently discussed~\cite{imw10}.

\noindent \underline{\it Conclusion}~:~
The utilitarian supersymmetric $U(1)_X$ gauge extension of the Standard Model 
of particle interactions proposed eight years ago~\cite{m02} allows for two 
classes of anomaly-free models which have no $\mu$ term and conserve baryon 
number and lepton number automatically.  A simple version with leptoquark 
superfields is discussed here with a number of interesting and verifiable 
properties.

The new $Z_X$ gauge boson of this model has specified couplings to quarks and 
leptons which are distinct from other gauge extensions and may be tested at 
the LHC.  The effective two-Higgs-doublet sector has $\tan \beta = 1$, and yet 
the mass of the lightest neutral Higgs boson may exceed the upper limit of 
126 GeV predicted in the MSSM.  A discrete $Z_4$ symmetry may be accommodated 
in the lepton sector so that the $3 \times 3$ neutrino mass matrix has two 
zero subdeterminants.  The scalar partners of the heavy singlet neutrinos 
could be components of dark matter, and since they may be decay products of 
the leptoquark fermions, they are possible sources of missing energy at the 
LHC.

\noindent \underline{\it Acknowledgement}~:~
This work was supported in part by the U.~S.~Department of Energy Grant 
No. DE-FG03-94ER40837.

\bibliographystyle{unsrt}

\end{document}